\documentclass[10pt]{iopart}
\usepackage{amsfonts}
\usepackage{graphicx}

\begin{document}

\title{The luminosity-redshift relation in brane-worlds:\\
II. Confrontation with experimental data }
\author{Gyula M. Szab\'{o}$^1$, L\'{a}szl\'{o} \'{A}. Gergely$^{1,2,3}$, Zolt\'{a}n
Keresztes$^{1,2}$ \\
\address: $^1$ Department of Experimental Physics, University of Szeged,
D\'om t\'er 9, Szeged 6720, Hungary \\
$^2$ Department of Theoretical Physics, University of Szeged,
Tisza Lajos k\"or\'ut 84-86, Szeged 6720, Hungary \\
$^3$ Department of Applied Science, London South Bank University,
103 Borough Road, London SE1 0AA, UK \\
\tiny{Email: Gyula M. Szab\'{o} - szgy@titan.physx.u-szeged.hu;
L\'{a}szl\'{o} \'{A}. Gergely - gergely@physx.u-szeged.hu;
Zolt\'{a}n Keresztes - zkeresztes@titan.physx.u-szeged.hu} 
}

\begin{abstract}
The luminosity distance - redshift relation for a wide class of generalized
Randall-Sundrum type II brane-world models with Weyl fluid is compared to
the presently available supernova data. We find that there is a class of
spacially flat models with different amounts of matter $\Omega _{\rho }$ and
Weyl fluid $\Omega _{d}$, which have a very similar fitting quality. The
best-fit models are equally likely and can be regarded as extensions of the $%
\Lambda $CDM model, which is also included. We examine three models with
different evolutionary history of the Weyl fluid, characterized by a
parameter $\alpha =0,~2$ and$~3$. The first model describes a brane which
had radiated energy into the bulk some time ago, but in recent times this
energy exchange has ceased and only a dark radiation ($\alpha =0$) is left.
In the other two models the Weyl-fluid describes a radiating brane
throughout the cosmological evolution, up to our days. We find that the
trought of the fitting surface extends over a wider $\Omega _{d}$-range with
increasing $\alpha $, but the linear correlation of $\Omega _{d}$ and $%
\Omega _{\rho }$ holds all over the examined $\Omega _{d}$ range.
\end{abstract}

\section{Introduction}

Current observational data \cite{SDSS k=0}-\cite{WMAP3y} suggest that the
cosmological model of a Universe with only baryonic matter has to be
modified. In the easiest way, the model can be reconciled with observations
by the introduction of a cosmological constant $\Lambda$ and of considerable
amount of dark matter ($\Lambda $CDM model). Because the energy densities of
both baryonic and dark matter decrease during cosmological evolution, the
cosmological constant will dominate the late-time evolution. This process
was first suggested for the explanation of Ia supernovae data, which suggest
that our Universe has reached an accelerating phase. In a $\Lambda$%
-dominated universe, the luminosity distance increases faster with redshift
than in the model without $\Lambda $ \cite{SahniStarobinski}, exactly as
required by the supernova data.

Generally, the agreement with experiments can be achieved by introducing a
dark energy component of the Universe, which replaces $\Lambda $. Such a
dark energy in general does not clump. A recent analysis \cite{TavakolFay}
shows that a dark energy model with varying dark energy density going
through a transition from an accelerating to a decelerating phase at
redshift $0.45$ fits well the observational data. Based on observations, the
dark energy equation of state $w=p/\rho $ is within about $-1\pm 0.1$ \cite%
{Krauss}.

It has been expected for some time that alternative gravitational theories,
motivated by string / M-theory could replace dark matter and dark energy by
geometric effects. The curved generalizations (see for example the review 
\cite{MaartensLR}) of the original Randall-Sundrum type II model \cite{RS2}
consist of a hypersurface with tension $\lambda $ (the brane), representing
our observable universe, embedded in a $5$-dimensional space-time (the
bulk). Gravitational dynamics on the brane is governed by an effective
Einstein equation \cite{SMS}, \cite{Decomp}. The sources of gravity in the
effective Einstein equation include terms due to the asymmetric embedding of
the brane into the bulk \cite{Decomp}, non-standard model fields in the
bulk, and even quantum corrections approximated as induced gravity effects 
\cite{DGP}-\cite{Induced}.{\ }

The most relevant source term for early cosmology is a quadratic source term
in the energy-momentum tensor \cite{BDEL}. This term dominates over the
linear term before the Big Bang Nucleosynthesis (BBN). In the simplest case
of cosmological symmetries and suppression of the energy exchange between
the brane and the bulk and whenever the bulk contains a static black hole,
the Weyl curvature of the bulk generates a so-called Weyl fluid effect on
the brane. In Fig 1 of our companion paper \cite{BraneLuminosityDistance1}
(to be referred in what follows as paper I) we classify the different
brane-world theories and their inter-relations. They are divided into two
branches, one containing the original Randall-Sundrum type II model (BRANE1)
and the other the flat DGP model (BRANE2). The model with Weyl fluid belongs
to the BRANE 2 branch.

Supernova data were confronted with the induced gravity models \cite{Goobar}-%
\cite{MM}. When they are combined with the Sloan Digital Sky Survey (SDSS)
baryonic peak, these seem to rule out the flat DGP models \cite{Goobar}, 
\cite{Sahni}. However it was argued in \cite{MM} that the Cosmic Microwave
Background (CMB) shift parameter can over-turn this conclusion. Structure
formation and CMB were also considered in the DGP models in Ref. \cite%
{RoyStrukt}.

Most recently, the authors of \cite{gao} tested the accelerating phase of
the universe's expansion with a comparison of the models and the supernova
data. They have tested the $\Lambda $CDM model, the DGP model and three $w$%
CDM models with equations of state where $w(a)$ (i) was constant with scale
factor $a$, (ii) varied as $w(a)=w_{0}+w_{a}(1-a)$ for redshifts probed by
the supernovae but fixed at $-1$ for earlier epochs, and (iii) varied as $%
w_{0}+W_{a}(1-a)$ since the recombination. Their main conclusion is that all
the five examined models explain equally well the acceleration, and none of
them could be selected as a preferred model, based on the Ia-type supernova
data.

The authors of Ref. \cite{Sahni} have compared the predictions of the flat
BRANE1 and BRANE2 models to the Gold \cite{gold} and Supernova Legacy Survey
(SNLS)\ \cite{SNLS} supernova data sets, incorporating the baryon acoustic
peaks into the analysis. These brane-world models in certain parameter range
(when their induced gravity parameter $\Omega _{l}$ is small; the flat DGP
models falling outside this range) are satisfied by both supernova data
sets. The BRANE1 models fit better to the SNLS data, while the BRANE2 models
fit better to the Gold data set. Since the analysis depends very weakly on
the bulk cosmological constant $\widetilde{\Lambda }$, the value of $%
\widetilde{\Lambda }$ was fixed at zero. With this modification, the BRANE1
model fits better to the SNLS data than the $\Lambda $CDM model and fits
comparably well to the Gold data. The same conclusion holds for the BRANE2
model. In the analysis of \cite{Sahni} the dark radiation dimensionless
parameter $\Omega _{d}$ is switched off.

Using two recent supernova data sets, the CMB shift parameter, and the
baryon oscillation peaks, the authors of Ref. \cite{LMM} have found that the
LDGP model (a subclass of the BRANE1 models with the effective energy
density having a phantom-like behavior due to extra-dimensional effects, see
Fig 1 of Paper I) fits the observations if it is very close to the $\Lambda $%
CDM model. The modification of the LDGP model with respect to the $\Lambda $%
CDM model appears in the form of a linear term in the Friedmann equation,
\thinspace $H/r_{c}$, where $H$ is the Hubble parameter and $r_{c}$ a
crossover scale. This model includes a cosmological constant, possibly
screened by the modified gravity, however the comparison with observations
sets strong constraints on the screening.

The first comprehensive study of the generalized Randall-Sundrum type II
(RS) brane-worlds tested against astronomical data was presented in Ref. 
\cite{DabrowskiBrane}. Agreement with earlier supernova data has been
established in the presence of a cosmological constant. In this analysis the
dark radiation from the bulk was switched off ($\Omega _{d}=0$) and the
energy-momentum squared term was kept. Under these assumptions, for flat
spatial sections and matter parameter $\Omega _{\rho }=0.3$ the maximum
likelihood method gave $\Omega _{\lambda }=0.004\pm 0.016$ for the parameter
characterizing the source term quadratic in the energy-momentum. This in
turn implies a tiny value of the brane tension, which is disfavored by
generic brane-world arguments. Moreover, much lower values for $\Omega
_{\lambda }$ emerge from both CMB and BBN.

In contrast to Refs. \cite{Sahni} and \cite{DabrowskiBrane} the analysis of 
\cite{Fay} keeps both $\Omega _{\lambda }$ and $\Omega _{d}$, the latter
obeying $\left\vert \Omega _{d}\right\vert <0.01$. The best fit is obtained
at $\Omega _{\rho }=0.15$, $\Omega _{\Lambda }=0.80\,$, $\Omega _{\lambda
}=0.026$ and $\Omega _{d}=0.008$. With the high value of the brane tension
set by either (a) the value of the $4$-dimensional Planck constant and
sub-millimeter tests \cite{tabletop} on possible deviations from Newton's
law (in units $c=1=\hbar $ these give $\lambda _{tabletop}^{\min }=138.59$
TeV$^{4}$ see \cite{GK}, \cite{MaartensLR}), (b) astrophysical
considerations $\lambda _{astro}^{\min }=5\,\times 10^{8}$ MeV$^{4}$ \cite%
{GM} or (c) BBN constraints $\lambda _{BBN}^{\min }=1$ MeV$^{4}$ \cite%
{nucleosynthesis}, the quadratic source term barely counts at late-times in
the cosmological evolution.

Given the high limits for the values of $\lambda $, in any realistic model $%
\Omega _{\lambda }$ can be safely ignored. This is a crucial difference of
our forthcoming analysis as compared to the one presented in Refs. \cite{Fay}
and \cite{DabrowskiBrane}, where the corresponding cosmological parameter $%
\Omega _{\lambda }$ was kept.

The next question is whether the source term arising from the Weyl curvature
of the bulk may be kept, in other worlds, whether $\Omega _{d}\neq 0$. The
Weyl curvature of the bulk gives an energy density $\rho _{d}=6m/\kappa
^{2}a^{4}$, where $\kappa ^{2}=8\pi G$ is the gravitational coupling
constant. In Ref. \cite{BBNLIM} it was shown that the BBN limits constrained
the dark radiation component as $-1.23\leq \rho _{d}\left( z_{BBN}\right)
/\rho _{\gamma }\left( z_{BBN}\right) \leq 0.11$. Combining this with CMB
constraints reduces this range to $-0.41\leq \rho _{d}\left( z_{BBN}\right)
/\rho _{\gamma }\left( z_{BBN}\right) \leq 0.105$. Here $\rho _{\gamma }$ is
the energy density of the background photons. Another constraint for the
value of the dark radiation at BBN was derived in \cite{Nucleosynthesys} as $%
-1<\rho _{d}\left( z_{BBN}\right) /\rho _{\nu }\left( z_{BBN}\right) <0.5$,
where $\rho _{\nu }$ is the energy density contributed by a single,
two-component massless neutrino. This constraint was derived for high values
of the $5$-dimensional Plank mass.

In the simplest case the Weyl source term evolves as a radiation, thus its
present value is obviously tiny. This is the reason why all mentioned
references \cite{Sahni} and \cite{DabrowskiBrane} comparing RS brane-worlds
with observations disregard dark radiation. But is this a necessary
assumption? Formulating the question the other way around: if we include
even a small component of dark radiation into the late-time universe model
we face a serious problem. Due to the fact that the energy density of dark
radiation decreases as $a^{-4}$ (compared to that of matter which is $a^{-3}$%
), even an amount of dark radiation of the same order as the amount of
baryonic matter nowadays implies dark radiation dominance in the past, for
example during structure formation. This conclusion is contradicted by
numerical simulations, which favorize cold dark matter as the dominant
component of the Universe during structure formation \cite{Millennium}.

However we can generalize the validity of the model by lifting the
requirement of a \textit{constant } mass $m$ in the dark radiation energy
density. A constant $m$ implies a static Schwarzschild-anti de Sitter bulk
and no energy exchange between the brane and the bulk. Therefore dark
radiation is a manifestation of an equilibrium configuration with a static
bulk, and it may be well possible that such a situation is reached only at
the latest stages of the evolution of the brane-world Universe. Whenever $m$
depends on a certain, non-zero power of $a$, the evolution of the energy
density of the Weyl source term evolves in a non-standard way, allowing to
escape from the argument of a small Weyl fluid left nowadays.

We propose here the LWRS (Lambda-Weyl fluid-Randall-Sundrum) model, a
specific \textit{RS model with i) cosmological constant, ii) the brane
radiating away energy during various stages of the cosmological evolution,
characterized by the index }$\alpha $ \textit{and iii) a Weyl fluid
depending on the actiul value of }$\alpha $ \textit{during the latest stage
of cosmological evolution}, which can be tested by supernova observations.
For the inclusion of the LWRS model in the classification of brane-world
models, see Fig 1 of paper I. \ 

The LWRS model takes into account the possibility of an energy exchange
between the brane and the bulk. This idea is not new. Indeed, it was already
proposed that during an inflationary phase on the brane radiation is emitted
and black holes thermally nucleate in the bulk \cite{ChKN}. Later on, but
still in the high energy regime, the brane radiates such that the \textit{%
mass function} of the bulk black hole increases with $a^{4}$ \cite{LSR}.
This means that the Weyl source term becomes a constant in this era. The
brane continues to radiate away energy during structure formation \cite%
{PalStructure}, a process leading to a bulk black hole mass function $%
m\propto a^{\alpha }$, with $1\leq \alpha \leq 4$. (Other models with the
brane radiating energy into the bulk are also known \cite{radiatingBrane}.)

For $\alpha =0\,$\ the Weyl fluid is known as dark radiation, for $\alpha
=2,~3$ it gives the correct growth factor during structure formation. For $%
\alpha =1,~4$ it is indistinguishable from dark matter and a cosmological
constant, respectively$.$ Therefore pure dark radiation can emerge only in
the low-$z$ limit, while at earlier times a dynamic bulk - brane interaction
governed by energy exchange should be present.

\section{Confronting the models with the selected supernova data}

As type Ia supernovae result from the explosion of white dwarf stars with
identical mass, they show remarkable similarities. By employing well
established calibration methods, one can calculate the maximal luminosity of
the object (in the reference system of the explosion). This is done by
analyzing the time-dependent variation of the emitted luminosity and the
spectrum, a method known as the Multi-Color Light Curve analysis \cite{Riess}%
, \cite{gold}. In this process the observed parameters, the \textit{shape\
of the light curve}\ and the \textit{spectral distribution of the emission}
have to be converted into the reference system of the host galaxy. For
distant supernovae this translates to take into account the time dilation
and the so-called $K$-correction \cite{Schmidt}. While these methods depend
on $z$, they are independent on the specific cosmological model. After
performing these corrections, we have well-calibrated maximal luminosities
for the supernovae of type Ia and in consequence they are considered as
standard candles.

In 2003 a list of $d_{L}$--$z$ data pairs were published for 230 supernovae
of type Ia \cite{Tonry}, and 60 of them had low absoprtion (i.e. $A_V.1$) and
cosmological redshift ($z>0.01$). To give result which can easily be compared to
earlier works, we also involve this selected low-absorption 
data set for the most of the examinations. The
basic supernova data we use here is the improved Gold set \cite{gold06},
which was released in 2006.

We confront with supernova observations several models from paper I. In Fig %
\ref{Fig3} we represent graphically on both logarithmic and linear scales
their luminosity distance - redshift relations up to $z=2.5$ . The plots are
for $k=0$ and$\ \Omega _{\rho }=0.27$ (according to the combined analysis of
the SDSS and WMAP 1-year data in Ref \cite{SDSS WMAP k=0}). In particular,
the luminosity distance - redshift relation is shown for the following
models:

\begin{itemize}
\item The LWRS model (the perturbative solution given by Eqs. (56),
(60)-(61), (63), (65) and (67) of paper I, with $\Omega _{\lambda }=0$ and
for $\alpha =0$) for the two values of the late-time dark radiation $\Omega
_{d}=-0.05$ and $\Omega _{d}=0.05$\ (the curves 1 and 3, respectively). The
latter models contain a brane which radiates energy at early times (for $%
\Omega _{d}>0$) and during structure formation, such that a bulk black hole
is formed and its mass increases continuously. As this process slows down,
the Weyl curvature of the bulk induces the late-time dark radiation on the
brane.

\item The $\Lambda $CDM model, given by Eqs. (60)-(61) of paper I (curve 2).

\item The solutions with brane tension $\lambda =2\Lambda /\kappa ^{2}$ and
no dark radiation (given by Eq. (52) of paper I) for both admissible values
for this model, at $\Omega _{\Lambda }=0.704$ (curve 4) and $\Omega
_{\Lambda }=0.026$ (curve 6). The former is similar to the class of models
discussed in \cite{DabrowskiBrane}.

\item The late-time universe $\Omega _{\lambda }=0$ limit of the RS model
with Randall-Sundrum fine-tuning, containing a huge amount of dark radiation 
$\Omega _{d}=0.73$, given by Eq. (44) of paper I (curve 5). 
\begin{figure}[tbp]
\includegraphics[bb=50 50 400 315, height=5.2cm]{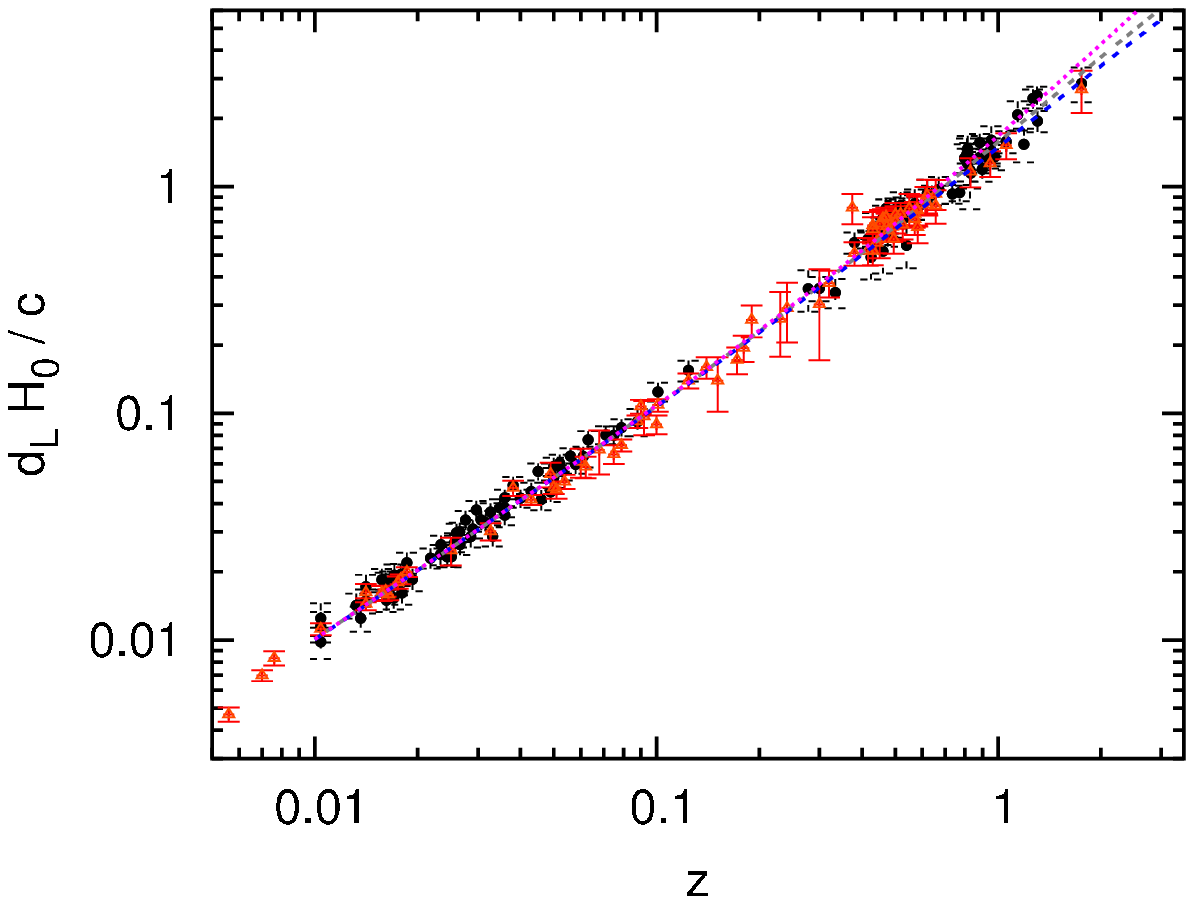} %
\includegraphics[bb=50 50 400 315, height=5.2cm]{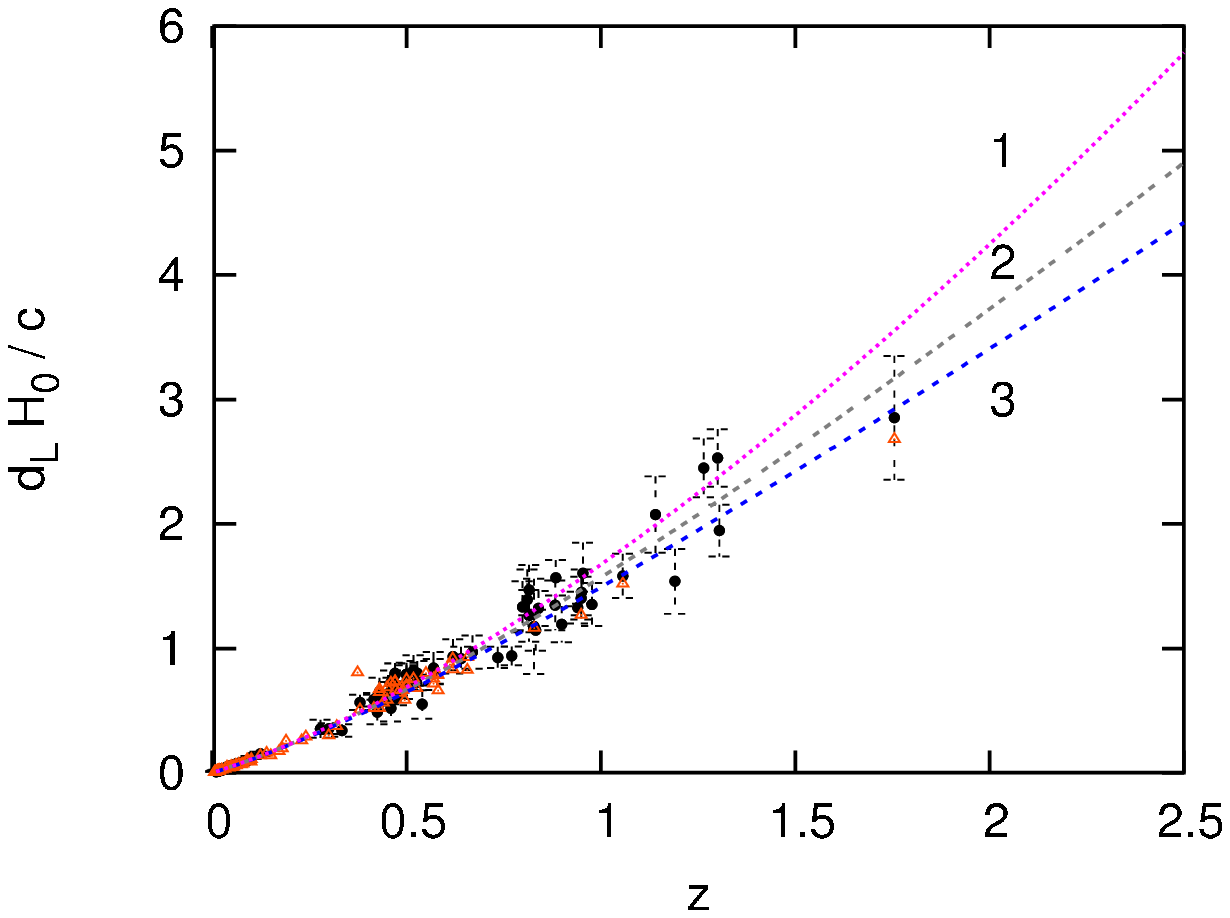} %
\includegraphics[bb=50 50 400 315, height=5.2cm]{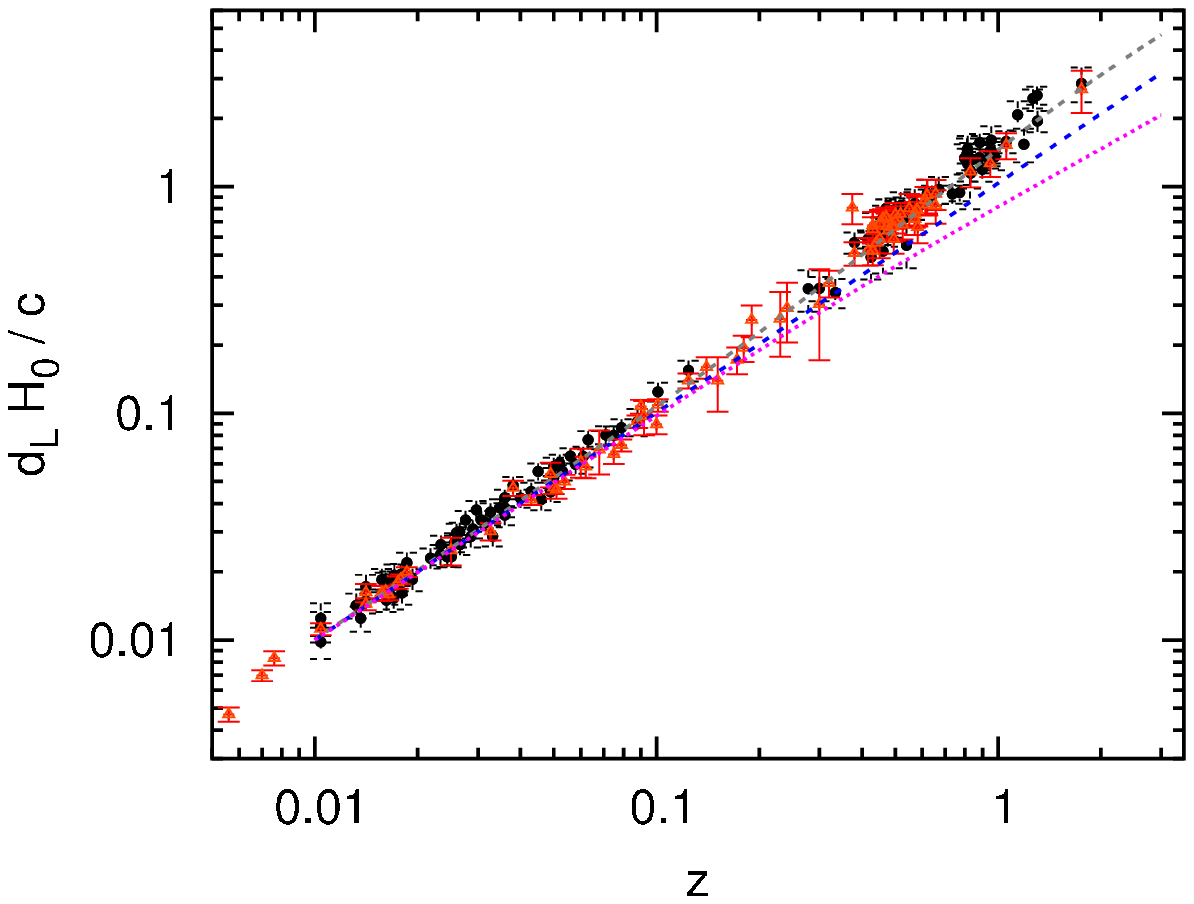} %
\includegraphics[bb=50 50 400 315, height=5.2cm]{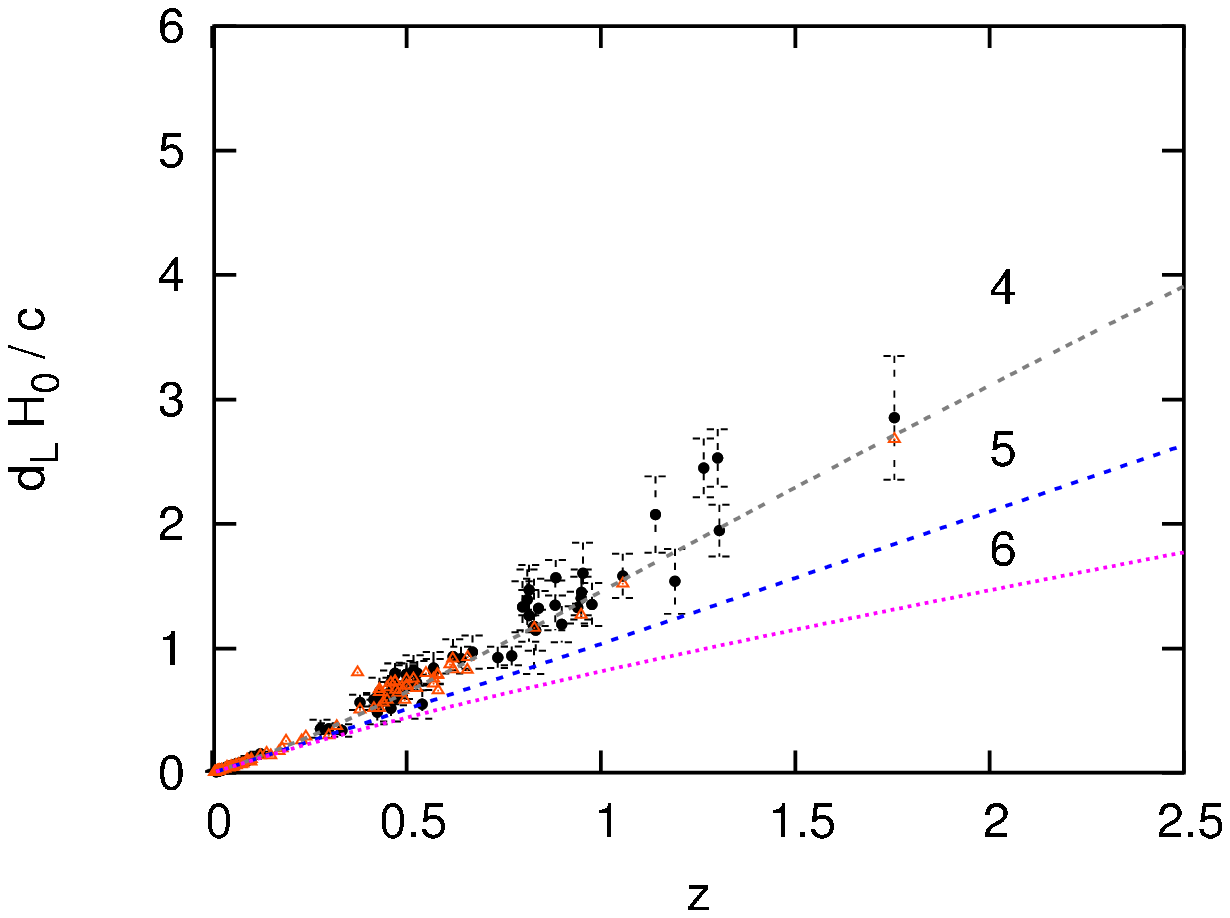}
\caption{(Color online) Luminosity distance -- redshift relations for
selected brane-world cosmologies and for the $\Lambda $CDM model, compared
to the supernova data. The diagrams are log-scaled (left panel) and linearly
scaled (right panel). Selected low absorption supernovae from Ref. 
\protect\cite{Tonry} are plotted with red, black dots represent the Gold set 
\protect\cite{gold}. Both sets are represented with the corresponding error
bars on the log-scaled diagrams. For the sake of perspicuity, the error bars
of low absorption supernovae are not represented on the linearly scaled
diagrams. The plotted models are the $\Lambda $CDM model (2); the brane
models with cosmological constant and late-time dark radiation (1 and 3);
without cosmological constant but with dark radiation (5); with cosmological
constant satisfying $\Lambda =\protect\kappa ^{2}\protect\lambda /2$, thus
low brane tension (4 and 6) and no dark radiation.}
\label{Fig3}
\end{figure}
\end{itemize}

In Fig. 1 we plot these models in a comparison to low-absorption supernova data from Ref. 
\cite{Tonry} (red triangles) together with the Gold set \cite{gold} (black
dots). The error bars are indicated in the respective colors. The diagrams
with linear scale are more instructive, as they emphasize the difference
among the predictions of the chosen models and how they fit data, while the
logarithmic scale better disseminate between the low $z$ points.

The models represented by the curves 1, 3 and 4 by eye seem to compare as
well with the supernova observations as the $\Lambda $CDM model (curve 2).
By contrast, the models represented by the curves 5 and 6 seem to be not
supported by observations. The model with no cosmological constant and
significant dark radiation $\Omega _{d}=0.73$, $\Omega _{\lambda }=0$ (curve
6) and the model with $\Lambda =\kappa ^{2}\lambda /2$ and $\Omega _{\Lambda
}=0.025$ (curve 5) are significantly inconsistent with the observations, as
they give $\chi ^{2}=213 $ and $395$, 
respectively\footnote{We mention here that we have also excluded several other models with
Randall-Sundrum fine-tuning (not shown on Fig \ref{Fig3}), which have either
a very low value of the brane tension or a significant dark radiation. For
example, the models with $\Omega _{d}=0.0258835$, $\Omega _{\lambda
}=0.70412 $ and $\Omega _{d}=0.70412$, $\Omega _{\lambda }=0.0258835$ gave $%
\chi ^{2}=246$ and $415$, respectively.}. 
All other models shown on 
Fig \ref{Fig3} are comparable with the supernova observations, as it was expected by
a simple glance.

The $\chi ^{2}=50$ value found for the $\Lambda =\kappa ^{2}\lambda /2$%
-model with $\Omega _{\Lambda }=0.74$ (curve 4) is slightly better than $%
\chi ^{2}$ found the $\Lambda $CDM model. However, as mentioned earlier, the
tiny brane tension $\lambda =38.375\times 10^{-60}$TeV$^{4}$, several order
of magnitudes lower than all existing lower limits rules out this model as
well.

The best fitting models are the models with brane cosmological constant; a
high value of the brane tension (leading to $\Omega _{\lambda }\approx 0$)
and a small contribution of dark radiation, $\Omega _{d}=\pm 0.05$ (the
curves 1 and 3). For $\Omega _{d}=-0.05$ we find $\chi ^{2}=65$, which is
still acceptable. For $\Omega _{d}=0.05$ we get $\chi ^{2}=49$.

Values of $\Omega _{d}$ between these limits are also admissible. It is
likely that by increasing $\Omega _{d}$ towards higher positive values, $%
\chi ^{2}$ remains compatible, however the accuracy of the perturbative
solution is deteriorated with increasing $\Omega _{d}$, therefore higher
orders in the expansion would be necessary to take into account.

\begin{figure}[tbp]
\includegraphics[bb=50 50 400 315, height=5.2cm]{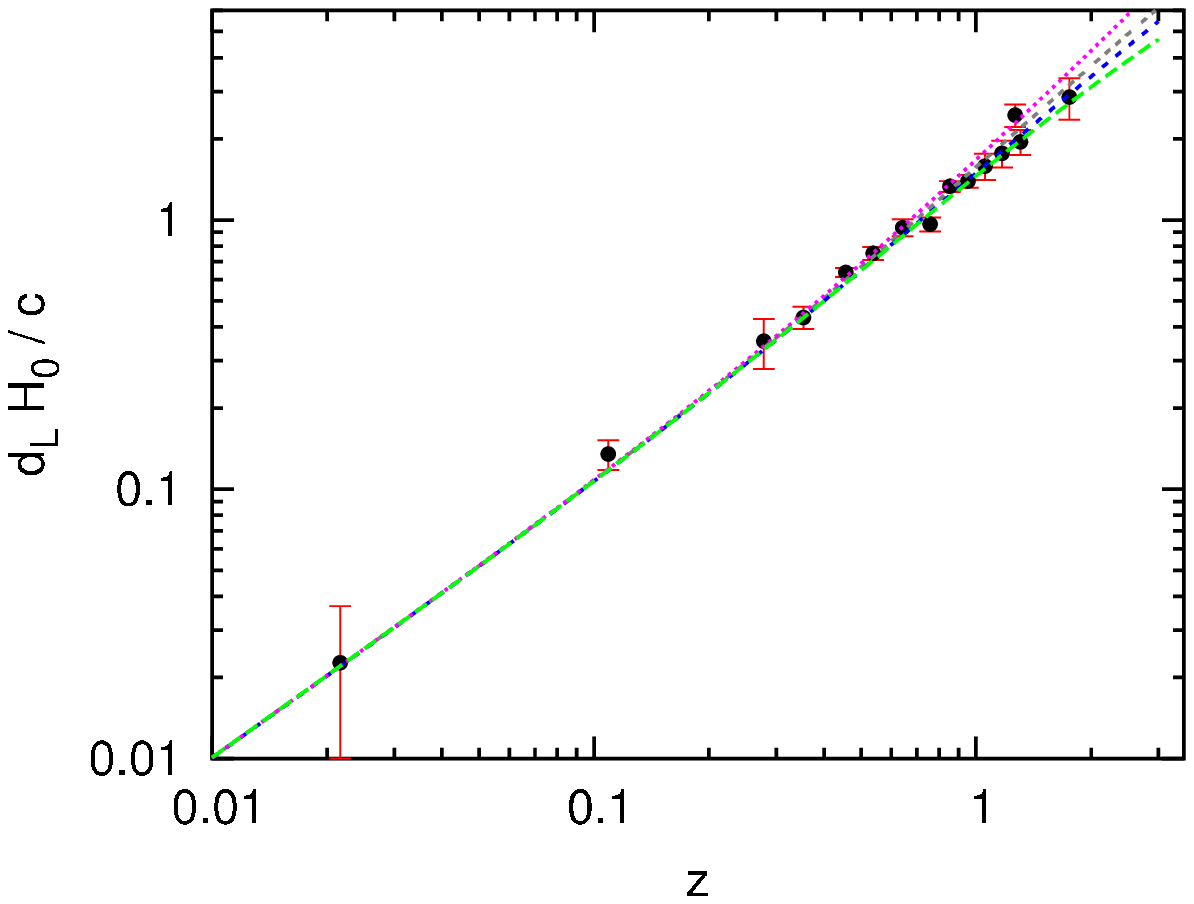} %
\includegraphics[bb=50 50 400 315, height=5.2cm]{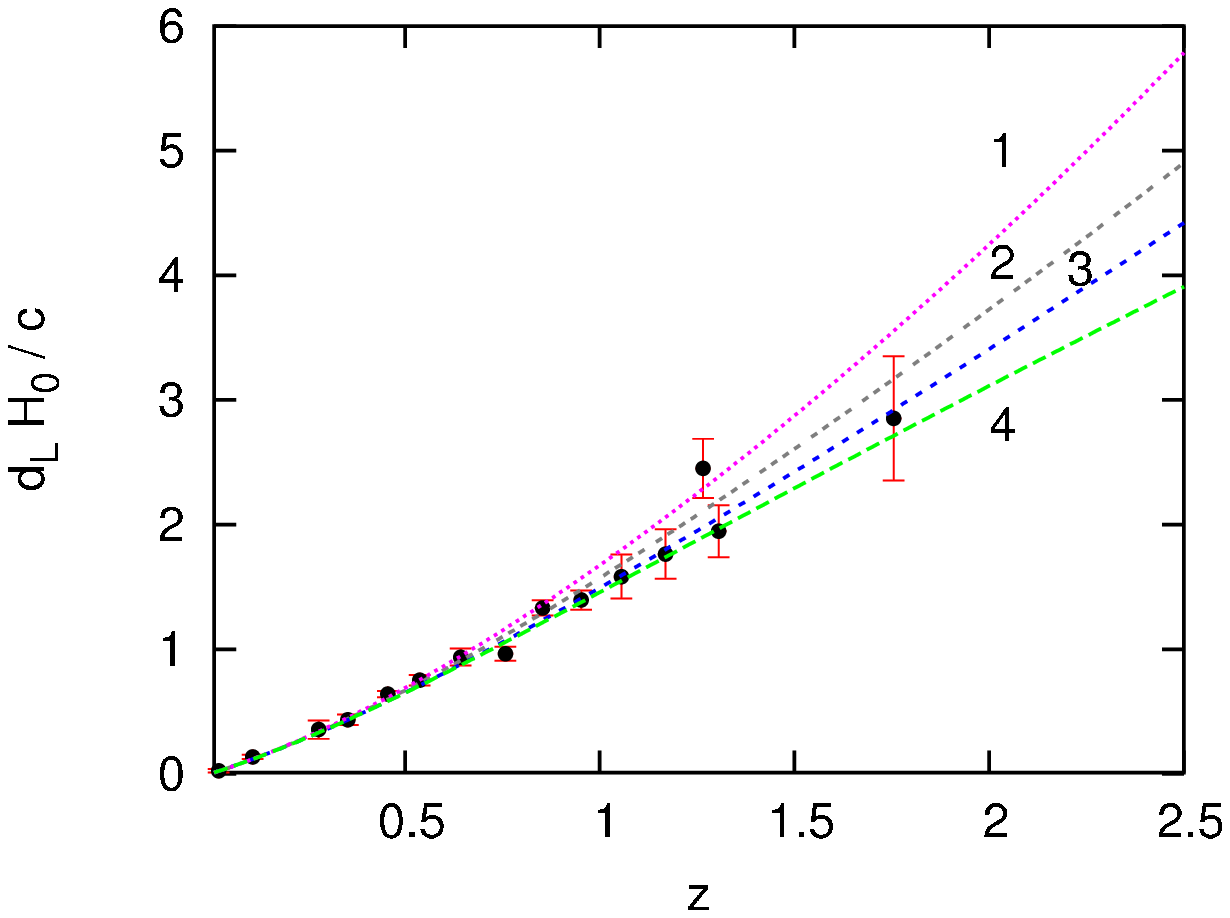}
\caption{(Color online) The luminosity distance - redshift relation for the
viable brane-world models and the $\Lambda $CDM model (the curves (1)-(4) of
Fig \protect\ref{Fig3}), both with logarithmic (left panel) and linear scale
(right panel), compared to the smeared Gold set \protect\cite{gold}. The
best fit is obtained for the brane-world model (3), with $5\%$ dark
radiation.}
\label{Fig4}
\end{figure}

\section{The Gold2006 set of supernovae}

More recently, Riess et al. \cite{gold06} have published a new set of $182$
gold supernovae, including new HST observations and recalibrations of the
previous measurements. It is an interesting question how this recalibration
influenced the above conclusions for the well-fitting models with dark
radiation.

We applied the same tests to the Gold2006 data set as described in the
previous section. First we assumed that $\Omega _{\rho }=0.27$, as before,
cf. Ref \cite{SDSS WMAP k=0}. In this case the critical value of $\chi ^{2}$
is $197$ at 80\%{} level and 209 at 90\%{} confidence level. 
Then the models represented by the curves 1-4 of Fig \ref{Fig4}
behave as follows. The model with a small amount of negative dark radiation
is disfavored at 80\%{} confidence ($\chi ^{2}=204$). 
The models with $\lambda=2\Lambda /\kappa ^{2}$ and $\Omega _{\Lambda }=0.704$ 
are ruled out at 90\%{} confidence level, too (as $\chi ^{2}=221$).
As expected from the previous analysis, the $\Lambda $CDM model ($\chi
^{2}=192$) and the LWRS model with $\Omega _{d}=0.05$ (giving $\chi ^{2}=194$%
) compete closely. We also remark that varying $\Omega _{d}$ between $-0.03$
and $0.07$, the $\chi ^{2}$ remains under the critical value.

\begin{figure}[tbp]
\includegraphics[height=8cm]{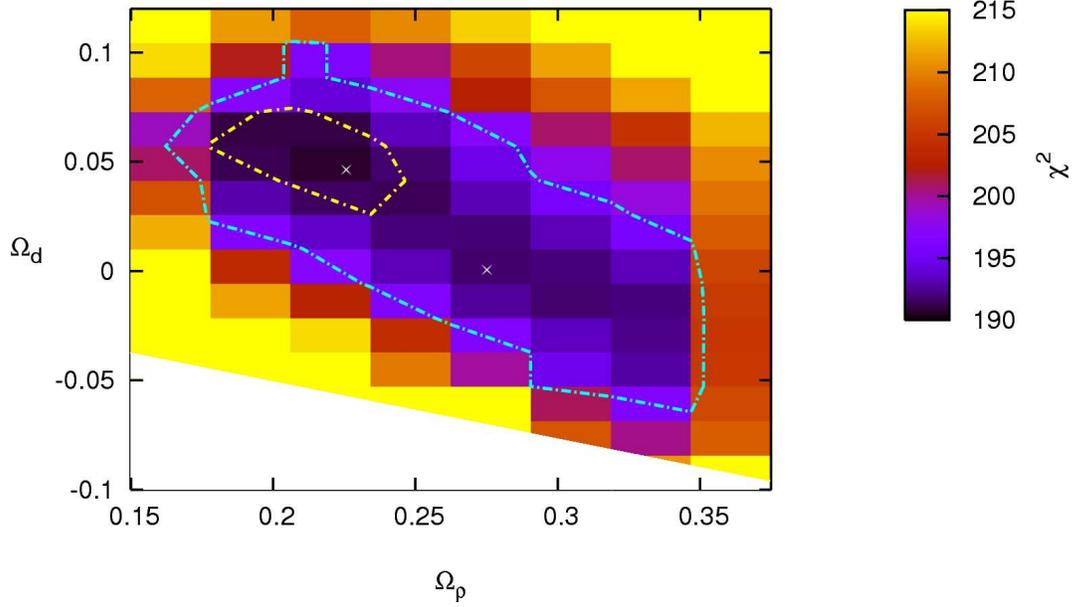}
\caption{(Color online) The fit of the luminosity distance - redshift
relation for the LWRS brane-world models with dark radiation, (including the 
$\Lambda $CDM model for $\Omega _{d}=0$). There is no assumption for $\Omega
_{\protect\rho }\in \left( 0.15,~0.35\right) $, its preferred value $0.225$
being determined from the supernova data, together with the preferred value $%
0.040$ of $\Omega _{d}$. The contours refer to the 1-$\protect\sigma $ and 2-%
$\protect\sigma $ confidence levels and both are centered on the LWRS model
with the values given above. The local minimum represented by the $\Lambda $%
CDM model is at $\Omega _{\protect\rho }=0.275$. Both the global and local
minima are marked. The white area in the lower left corner represents the
forbidden region of the parameter space for $z=3$. }
\label{Fig5}
\end{figure}

For gaining a deeper insight we have then calculated the predictions of the
models between $\Omega _{d}=-0.10\div 0.10$ with a stepsize of $0.01$ in $%
\Omega _{d}$, with $\Omega _{\rho }$ allowed to freely vary in the domain $%
0.15\div 0.35$ and $z$ in the range $0\div 3$. Then we looked for the best
fit of the Gold2006 set in the $\Omega _{d}-\Omega _{\rho }$ space. This is
represented on Fig \ref{Fig5}. The global minimum of the surface is at $%
\Omega _{d}=0.040$, $\Omega _{\rho }=0.225$ ($\chi ^{2}=190.52$), which
suggests an interesting opportunity for a Universe with less baryonic
density and with dark radiation, compatible with the Gold2006 supernova
data. The 1-$\sigma $ confidence interval is centered about this value. The $%
\Lambda $CDM model (where $\Omega _{d}$ is exactly $0$) has the local
minimum of $\Omega _{\rho }=0.275$ ($\chi ^{2}=195.8$), but this is outside
the 1-$\sigma $ confidence interval.

\begin{figure}[tbp]
\includegraphics[height=8cm]{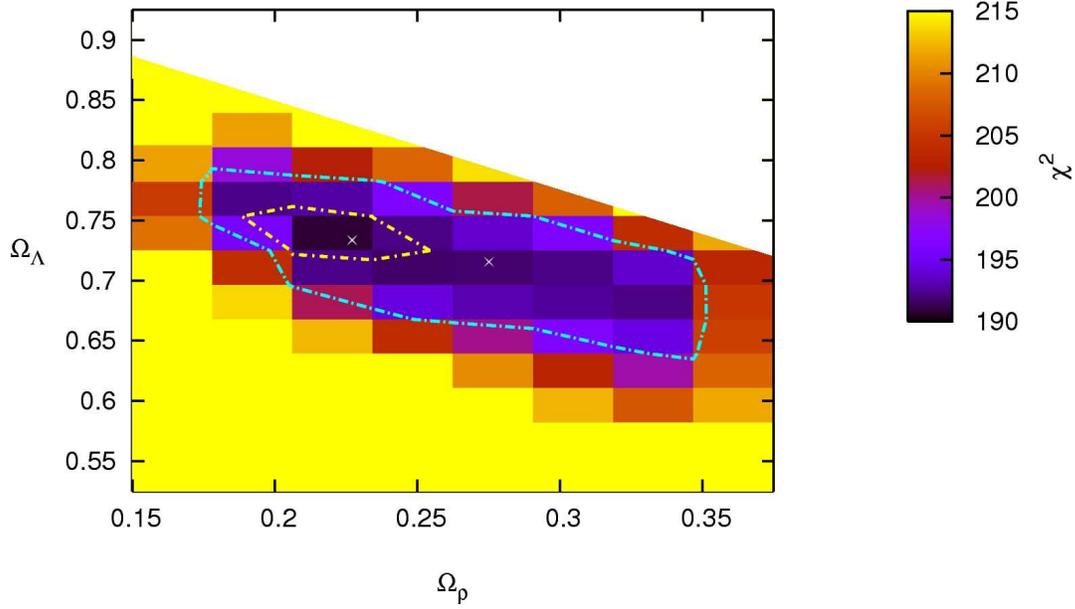}
\caption{(Color online) Same as on Fig \protect\ref{Fig5}, but in the $%
\Omega _{\Lambda }-\Omega _{\protect\rho }$ plane. The global minimum is at $%
\Omega _{\Lambda }=0.735$, $\Omega _{\protect\rho }=0.225$, while the local
minimum for the $\Lambda $CDM model gives $\Omega _{\Lambda }=0.725$ and $%
\Omega _{m}=0.275$ (both marked). The white area on the top right corner
represents the forbidden parameter range.}
\label{Fig6}
\end{figure}

Similar conclusions emerge from the plot in the $\Omega _{\Lambda }-\Omega
_{\rho }$ plane, Fig \ref{Fig6}. Here the global minimum of the surface is
at $\Omega _{\Lambda }=0.735$, $\Omega _{\rho }=0.225$. The local minimum of
the $\Lambda $CDM model is at $\Omega _{\Lambda }=0.725$.

We note that there is a forbidden parameter range in both planes $\Omega
_{d}-\Omega _{\rho }$ and $\Omega _{\Lambda }-\Omega _{\rho }$, represented
by white regions on Figs \ref{Fig5} and \ref{Fig6}. This is because the
Friedmann equation for these brane-world models%
\begin{equation}
\left[ \frac{H\left( z\right) }{H_{0}}\right] ^{2}=\Omega _{\Lambda }+\Omega
_{\rho }\left( 1+z\right) ^{3}+\Omega _{d}\left( 1+z\right) ^{4}>0\mathrm{{\ 
},}
\end{equation}
combined with $\Omega _{\Lambda }+\Omega _{\rho }+\Omega _{d}=1$ gives the
constraints%
\begin{equation}
\Omega _{d}\left[ \left( 1+z\right) ^{4}-1\right] +\Omega _{\rho }\left[
\left( 1+z\right) ^{3}-1\right] +1>0\mathrm{\ }
\end{equation}%
in the $\Omega _{d}-\Omega _{\rho }$ plane and%
\begin{equation}
\Omega _{\Lambda }\left[ \left( 1+z\right) ^{4}-1\right] -\left( 1+z\right)
^{3}\left[ 1+z\left( 1-\Omega _{\rho }\right) \right] <0\mathrm{\ }
\end{equation}%
in the $\Omega _{\Lambda }-\Omega _{\rho }$ plane.

The forbidden region increases in both cases with $z$. If we would like to
extend the limits to $z\rightarrow \infty $, we obtain the limiting curves $%
\lim_{z\rightarrow \infty }\Omega _{d}^{\min }\left( z,\Omega _{\rho
}\right) =0$ in the $\Omega _{d}-\Omega _{\rho }$ plane and $%
\lim_{z\rightarrow \infty }\Omega _{\Lambda }^{\max }\left( z,\Omega _{\rho
}\right) =1-\Omega _{\rho }$ in the $\Omega _{\Lambda }-\Omega _{\rho }$
plane. However the LWRS model being valid only for low values of $z$, we
represent on the graphs only the forbidden range for $z=3$.

\section{The compatibility of the LWRS model $\Omega _{d}=0.04$ and $\protect%
\alpha =0$ with cosmological evolution}

The energy density of dark radiation decreases too fast during cosmological
evolution to result in a considerable amount nowadays. We discuss this
problem and its possible remedy in detail here. First we comment on the
problem, then we show how an energy exchange between the brane and the bulk
can leave a considerable amount of dark radiation.

The constraint derived in \cite{BBNLIM} for the energy density of the dark
radiation: 
\begin{equation}
-0.41\leq \frac{\rho _{d}\left( z_{BBN}\right) }{\rho _{\gamma }\left(
z_{BBN}\right) }\leq 0.105\mathrm{,}  \label{BBNconstr}
\end{equation}%
where $\rho _{\gamma }\left( z_{BBN}\right) =\beta T_{BBN}^{4}$ \ is the
energy density of the background photons at the beginning of BBN. The
coefficient 
\begin{equation}
\beta =\frac{\pi ^{2}}{30}g_{\ast }\frac{k_{B}^{4}}{\left( \hbar c\right)
^{3}}=3.78\times 10^{-16}~g_{\ast }~\mathrm{{J~m}^{-3}~{K}^{-4}}
\end{equation}%
contains \cite{BDEL}, \cite{Garcia} the effective number $g_{\ast }$ of
relativistic degrees of freedom, which depends on the temperature. According
to \cite{KolbTurner} $g_{\ast }=10.75$ at the beginning of BBN, when $%
T_{BBN}=1.16\times 10^{10}~$K. Thus $\rho _{\gamma }\left( z_{BBN}\right)
=7.37\times 10^{25}~$J m$^{-3}$ emerges, giving the constraint%
\begin{equation}
-3.02\times 10^{25}~\mathrm{{Jm}^{-3}}\leq \rho _{d}\left( z_{BBN}\right)
\leq 7.74\times 10^{24}\mathrm{~{Jm}^{-3}.}  \label{constr}
\end{equation}%
Note, that the domain of allowable negative values is larger than the one
for positive values.

As for today the background photons have cooled to $T_{0}=2.725~$K and for
such low temperatures $g_{\ast }=3.36$\ \cite{Garcia}, \cite{KolbTurner}\
their energy density $\rho _{\gamma }=\rho _{\gamma }\left( z=0\right) $ is%
\begin{equation}
\rho _{\gamma }=7.01\times 10^{-14}\mathrm{{\ J m}^{-3}{\ }.}
\end{equation}%
With the value $H_{0}=73_{-3}^{+3}$ km~s$^{-1}$ Mpc$^{-1}$ of the Hubble
constant \cite{WMAP3y}, cf. Eq. (23) of paper I the present day cosmological
parameters $\rho $ and $\Omega \,$\ (both for background and dark radiation)
relate as%
\begin{equation}
\rho _{d,\gamma }=9.00\times 10^{-10}~\Omega _{d,\gamma }\mathrm{{\ J m}^{-3}%
{\ }.}  \label{rhom}
\end{equation}%
Thus the present value of $\Omega _{\gamma }$ is 
\begin{equation}
\Omega _{\gamma }=7.74\times 10^{-5}\mathrm{{\ },}
\end{equation}%
which is quite negligible. If the Weyl source term were to evolve as
radiation, its value would be even smaller, cf. Eq. (\ref{BBNconstr}).
Indeed Eqs. (\ref{constr}) and (\ref{rhom}) imply 
\begin{equation}
-1.02\times 10^{-4}\leq \Omega _{d}\leq 2.62\times 10^{-5}\mathrm{{\ }.}
\label{Omegad1}
\end{equation}%
$\left\vert \Omega _{d}\right\vert $ is of the same order of magnitude or
smaller as $\Omega _{\gamma }$.

However if the brane is radiating during structure formation, the mass
parameter $m$ becomes a function of the scale factor $m\propto a^{\alpha } $%
, with $1\leq \alpha \leq 4$ \cite{PalStructure}. Then the energy density
scales as \thinspace $a^{4-\alpha }$.

Now let us suppose that the brane is in an equilibrium (non-radiating)
configuration with $\alpha =0$ in the domain $0\leq z\leq z_{1}.$ In a
preceding era $z_{1}<z\leq z_{\ast }$ the brane radiates such that $\alpha
\neq 0$, finally right after the beginning of BBN, at $z_{\ast }<z\leq
z_{BBN}$ there is equilibrium once more ($\alpha =0$). Here $z_{BBN}=\left(
T_{BBN}/T_{0}\right) -1=4.26\times 10^{9}.$ According to this evolution%
\begin{eqnarray}
\rho _{d}\left( z_{BBN}\right) &=&\rho _{d}\left( \frac{a_{0}}{a_{1}}\right)
^{4}\left( \frac{a_{1}}{a_{\ast }}\right) ^{4-\alpha }\left( \frac{a_{\ast }%
}{a_{BBN}}\right) ^{4}  \nonumber \\
&=&\rho _{d}\left( \frac{1+z_{1}}{1+z_{\ast }}\right) ^{\alpha }\left(
1+z_{BBN}\right) ^{4}.
\end{eqnarray}%
Inserting this in Eq. (\ref{constr}) and employing Eq. (\ref{rhom}) we
obtain:%
\begin{equation}
-1.\,\allowbreak 02\times 10^{-4}~\leq \left( \frac{1+z_{1}}{1+z_{\ast }}%
\right) ^{\alpha }~\Omega _{d}\leq \allowbreak 2.\,\allowbreak 62\times
10^{-5}\mathrm{~.}
\end{equation}%
In the particular case $\alpha =0$ we recover the constraint (\ref{Omegad1})
set on pure dark radiation. However for any $\alpha >0$ we get

\begin{equation}
z_{\ast }\geq \left( 1+z_{1}\right) ~\left[ \max \left( -0.98~\Omega
_{d},~3.\,\allowbreak 82~\Omega _{d}\right) \right] ^{1/\alpha }\times
10^{4/\alpha }-1\mathrm{{\ }.}
\end{equation}

Let us specify this result for the best fit value $\Omega _{d}=0.04$.
Depending on $\alpha $ we obtain the following numerical relations between
the redshifts characterizing the switching on and off of the radiation
leaving the brane: 
\begin{equation}
z_{\ast }\geq \left\{ 
\begin{array}{cc}
1527.\,\allowbreak 80+1528.\,\allowbreak 80~z_{1} & ,\qquad \alpha =1 \\ 
38.\,\allowbreak 10+39.\,\allowbreak 10~z_{1} & ,\qquad \alpha =2 \\ 
\allowbreak 10.\,\allowbreak 52+11.\,\allowbreak 52~z_{1} & ,\qquad \alpha =3
\\ 
5.\,\allowbreak 25+6.\,\allowbreak 25~z_{1} & ,\qquad \alpha =4%
\end{array}%
\right. ~.
\end{equation}%
It is evident that the value of $z_{\ast }$ increases with $z_{1}$ (this
dependence becoming an approximate scaling for higher values of $z_{1}$) and
decreases with $\alpha $.

The lower limit in the LWRS model is $z_{1}=3$. Then%
\begin{equation}
z_{\ast }\geq \left\{ 
\begin{array}{cc}
6114.\,\allowbreak 20 & ,\qquad \alpha =1 \\ 
155.\,\allowbreak 40 & ,\qquad \alpha =2 \\ 
\allowbreak 45.\,\allowbreak 08 & ,\qquad \alpha =3 \\ 
24.\,\allowbreak 01 & ,\qquad \alpha =4%
\end{array}%
\right. ~.
\end{equation}
For the higher values of $\alpha $ the duration of the radiative brane
regime necessary to produce a high value of $\Omega _{d}$ today is quite
short.

\section{LWRS models with $\protect\alpha =2$, $3$ confronted with supernova
data}

In the absence of a known mechanism for changing $\alpha $, we examine here
the cases when $\alpha =2$ and $\alpha =3$ hold throughout the cosmological
evolution, up to nowadays. For this we confront these models with the
Gold~2006 set exactly as described before. To preserve the validity of the
perturbative solution, the range of $\Omega _{d}$ was selected to be $-0.1$--%
$0.1$, and we probed the range $0.15$--$0.35$ of $\Omega _{\rho }$. The
assumption for flatness was kept, too.

The results are qualitatively similar to the $\alpha =0$ case. The
remarkable difference is that the peak of the minimum turned into a
\textquotedblleft trough\textquotedblright , which lies aslope in the $%
\Omega _{\rho }$--$\Omega _{d}$ space. This means that instead of a district
solution, a complete model family exists in both cases, which can equally
well explain the supernova data. The $\Omega _{\rho }$ dependence of $\Omega
_{d}$ is less in the $\alpha =2$ model as compared to $\alpha =0$, and is
very small if $\alpha =3$. The $\Omega _{d}=0$ case is the $\Lambda $CDM
model where these model intersect. The steeper slope of the minimum trough
thus allows a much lower range of $\Omega _{\rho }$ in the $\alpha =2$, and
especially in the $\alpha =3$ models, with a value close to $0.3$. On the
other hand, the range of $\Omega _{d}$ gets more and more wide with
increasing $\alpha $, which results in the conclusion that the presence of $%
\Omega _{d}$ is mathematically plausible, and they have to be accounted for
in RS cosmology.

Due to the higher slopes of the 1-$\sigma $ and 2-$\sigma $ contours in
these $\alpha =2,~3$ models, $\Omega _{\rho }$ is much less affected by the
Weyl fluid, while $\Omega _{d}$ can have various values in the detriment of $%
\Omega _{\Lambda }$. Therefore the Weyl fluid can explain some of the dark
energy.

\begin{figure}[tbp]
\includegraphics[bb=142 163 490 750, height=12cm, angle=270]{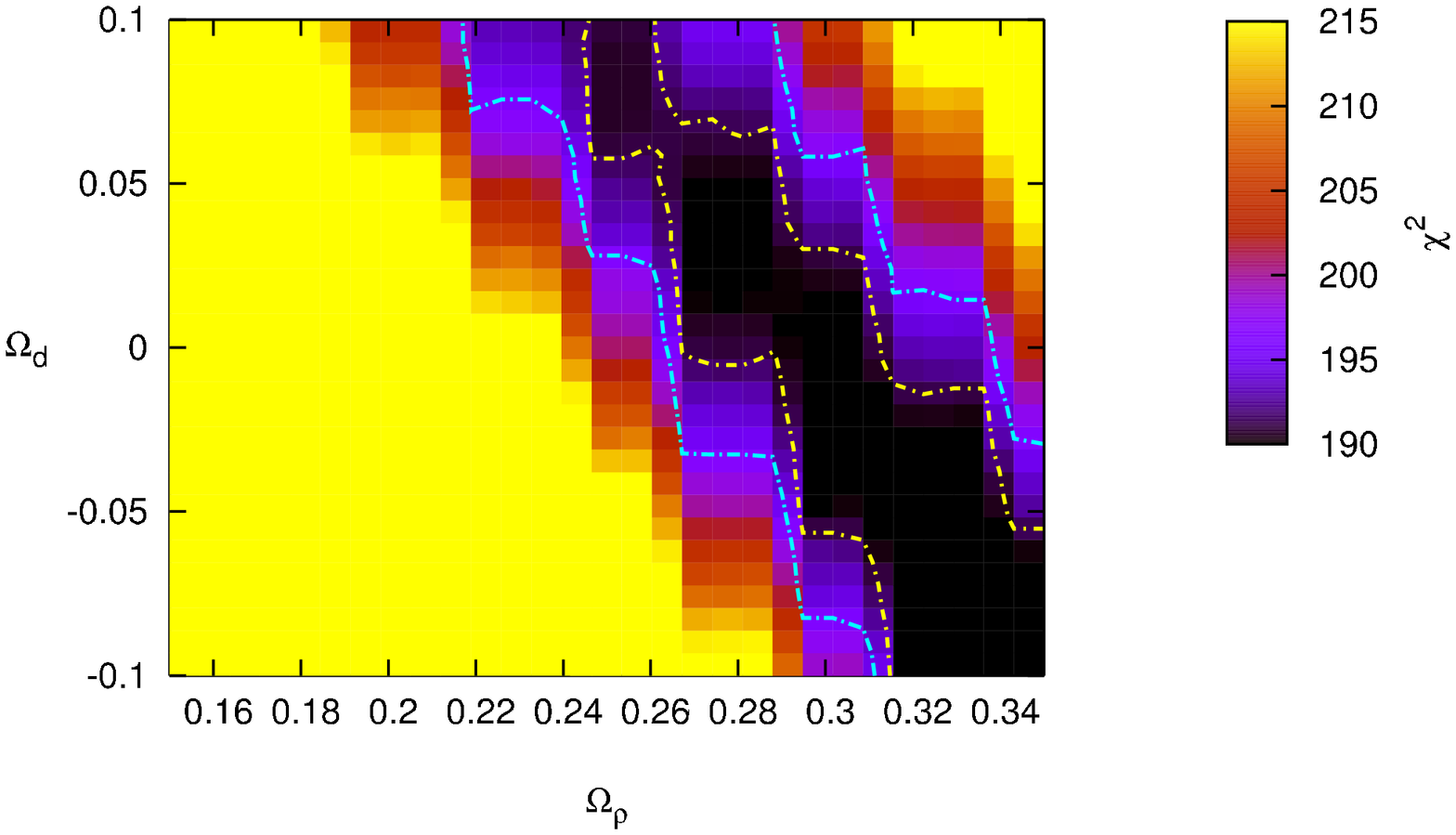}
\caption{(Color online) Same as on Fig \protect\ref{Fig5}, but for the $%
\protect\alpha =2$ models.}
\label{Fig7}
\end{figure}

\begin{figure}[tbp]
\includegraphics[bb=142 163 490 750,height=12cm, angle=270]{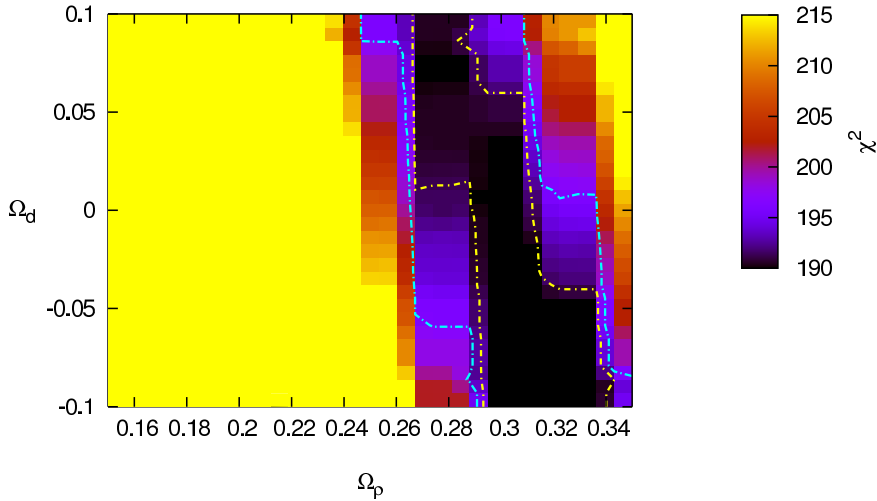}
\caption{(Color online) Same as on Fig \protect\ref{Fig5}, but for the $%
\protect\alpha =3$ models.}
\label{Fig8}
\end{figure}

\section{Conclusions}

The luminosity distance given in paper I as function of redshift in terms of
elementary functions and elliptical integrals of first and second type for
various brane-world models with Weyl fluid was confronted with the available
supernova data sets, including the Gold2006 data \cite{gold06}. The tested
models were:

(A) The models with Randall-Sundrum fine-tuning, discussed in section 4 of
paper I, with a considerable amount of dark radiation as a bulk effect, and
a high value of the brane tension.

(B) The two models discussed in subsection 5.1 of paper I, which obey $%
\Lambda =\kappa ^{2}\lambda /2$, have no dark radiation and were integrable
in terms of elementary functions.

(C) The LWRS models (subsection 5.2 of paper I), with a brane cosmological
constant, for which the luminosity distance could be given analytically as
function of redshift to first order accuracy in the dark radiation. (Due to
its smallness, the source term $\Omega _{\lambda }$ quadratic in the energy
density was suppressed in the perturbative models of paper I.)

The brane-world models (A) although interesting for historical reasons, do
not comply with observations. Even if we introduce an extremely high amount
of dark radiation $\Omega _{d}=0.73$, tentatively replacing the cosmological
constant in the energy balance $\Omega _{\Lambda }+\Omega _{\rho }+\Omega
_{d}+\Omega _{\lambda }=1$, these models are quickly outruled by supernova
data (curve 5 of Fig \ref{Fig3}). Dark radiation is not capable to replace
the cosmological constant in producing a late-time acceleration, since it
scales as usual radiation. The more we go back in the past, the higher
becomes its domination over matter. Therefore a cosmological constant or
dark energy is still needed in the generalized Randall-Sundrum type II
models.

Our analysis has also dismissed immediately the model (B) with $\Omega
_{\Lambda }=0.026$. Surprisingly, the other toy model (B) with $\Omega
_{\Lambda }=0.704$ was in good agreement with the Gold2006 data, but ruled
out by its low value of the brane tension, similarly as the models discussed
in Ref. \cite{DabrowskiBrane}. A low brane tension is in disagreement with
various upper limits set by cosmological and astrophysical tests.

The perturbative approach of subsection 5.2 of paper I can be considered
valid for a Weyl fluid with $-0.1<\Omega _{d}<0.1$. In this range the LWRS
brane-world models (C) were confronted with supernova data and for $\alpha
=0 $ the dark radiation with significant negative energy density ruled out.
The fact that a positive dark radiation (corresponding to a bulk black hole
rather than to a bulk naked singularity) is favoured by the presently
available best supernova data is in accordance with the early behavior of
the RS model with late-time dark radiation, where the brane radiating away
energy in early times leads to a black hole, which can further grow during
structure formation.

The remaining LWRS brane-world models with $\alpha =0$ and $\Omega _{d}$
between $-0.03$ and $0.07$ (and $\Omega _{\Lambda }$ changed accordingly)
turned out to be excellent candidates for describing our universe, as they
show remarkable agreement with the Gold2006 supernova data sets. If $\Omega
_{\rho }$ is allowed to vary in the range $\left( 0.15,~0.35\right) $, the
preferred values are $\Omega _{d}=0.040$, $\Omega _{\rho }=0.225,~\Omega
_{\Lambda }=0.735$.

The preferred cosmological parameters determined by comparing the LWRS model
with $\alpha =0$ with supernova data alone are in perfect accordance with
the WMAP 3-year data. Indeed according to Ref. \cite{WMAP3y} $\Omega _{\rho
}h^{2}=0.127_{-0.013}^{+0.007}$ and $h=0.73_{-0.03}^{+0.03}$ from which $%
\Omega _{\rho }=0.238_{-0.041}^{+0.035}$ emerge. The value of $\Omega _{\rho
}$ determined by comparing the LWRS model with the supernova data alone is
well in the middle of the domain allowed by the WMAP 3 year data.

We have then proved that the preferred value of $\Omega _{d}=0.04$ is
compatible with the known history of the Universe if the brane radiates away
energy into the bulk during a relatively short period of the cosmological
evolution. Such a process occurring between $z=24$ and $z=3$ could increase
the amount of dark energy today with a factor of $10^{3}$ as compared to the
non-radiating brane, exactly as required by the LWRS model with $\alpha =0$.

The LWRS models with $\alpha =1\ $($\alpha =4$)\ are identical with the $%
\Lambda $CDM model with the only difference that some fraction of the dark
matter (of the cosmological constant) has geometric origin.

Finally, the LWRS models with $\alpha =2\ $and $\alpha =3$\ do not present a
sharp minimum, but rather an elongated trought shape in the parameter space,
with the slope increasing with the value of $\alpha $ and $\Omega _{\rho
}\approx 0.3$. This means that in this class of models a wide range of
values for $\Omega _{d}$ (with a slight preference for negative values) and
corresponding values for $\Omega _{\Lambda }$ are fitting to the supernova
data.

We must note that the reliability of these values is somehow deteriorated by
the relatively small number of high-$z$ supernova and by the inherent
difficulties in the calibration of the available data. An obvious
source of error is that data from the Gold2006 set is a combination of
measurements taken on different instruments \cite{Tao} and in fact it has
been already signaled that the Gold2006 data set is not statistically
homogeneous \cite{Perivolaropoulos}.

The conclusion of this paper is somewhat similar to that of Ref. \cite{gao}:
the presently available supernova data are not enough to discern among
several cosmological models. However the difference between the predictions
of the acceptable models of our analysis (the $\Lambda $CDM model, the LWRS
brane-world with $\alpha =0$ and $\Omega _{d}=0.04\,\ $and the models with
Weyl fluid and $\alpha =2,~3$) are increasing with $z$. One may reasonably
hope that the very far ($z>2$) supernovae, which will be discovered for sure
in the following decade, will improve their comparison.

\ack 
We thank Botond Nagy for interactions in the early stages of this research.
This work was supported by OTKA grants no. 46939, 44665, 42509 and 69036. 
L\'{A}G and GyMSz were further supported by the J\'{a}nos Bolyai Grant of the
Hungarian Academy of Sciences and GyMSz by the Magyary Zolt\'{a}n Higher
Educational Public Foundation.


\section*{References}

\begin{thebibliography}{99}
\bibitem{SDSS k=0} Doroshkevich A, Tucker D L, Allam S and Way M J, \textit{%
Large scale structure in the SDSS galaxy survey }2004 \textit{Astron.
Astrophys.} \textbf{418} 7

\bibitem{SDSS WMAP k=0} Tegmark M, Strauss M A, Blanton M R et al., \textit{%
Cosmological parameters from SDSS and WMAP} 2004 \textit{Phys. Rev.} D 
\textbf{69} 103501

\bibitem{WMAP3y} Spergel D N, Bean R, Dor\'{e} O et al., \textit{Wilkinson
Microwave Anisotropy Probe (WMAP) Three Year Results: Implications for
Cosmology} 2006 \textit{astro-ph/0603449}

\bibitem{SahniStarobinski} Sahni V and Starobinski A, \textit{The Case for a
Positive Cosmological Lambda-term} 2000 \textit{Int. J. Mod. Phys.} D 
\textbf{9} 373

\bibitem{TavakolFay} Fay S and Tavakol R, \textit{A model-independent dark
energy reconstraction schem using the geometrical form of the
luminosity-distance relation }2006 \textit{Phys. Rev.} D \textbf{74} 083513

\bibitem{Krauss} Krauss L M, Jones-Smith K, Huterer D, \textit{Dark Energy,
A Cosmological Constant, and Type Ia Supernovae} 2007 \textit{%
astro-ph/0701692}

\bibitem{MaartensLR} Maartens R, \textit{Brane-world Gravity} 2004\textit{\
Living Rev. Rel}. \textbf{7} 1

\bibitem{RS2} Randall L and Sundrum R, \textit{An Alternative to
Compactification} 1999 \textit{Phys. Rev. Lett.} \textbf{83} 4690

\bibitem{SMS} Shiromizu T, Maeda K and Sasaki M, \textit{The Einstein
Equations on the 3-Brane World} 2000 \textit{Phys. Rev.} D \textbf{62} 024012

\bibitem{Decomp} Gergely L \'{A}, \textit{Generalized Friedmann branes} 2003 
\textit{Phys. Rev.} D \textbf{68} 124011

\bibitem{BDEL} Bin\'{e}truy P, Deffayet C, Ellwanger U and Langlois D, 
\textit{Brane cosmological evolution in a bulk with cosmological constant}
2000 \textit{Phys. Lett. }B \textbf{477} 285

\bibitem{DGP} Dvali G, Gabadadze G and Porrati M, \textit{4D Gravity on a
Brane in 5D Minkowski Space} 2000 \textit{Phys. Lett. }B\textit{\ }\textbf{%
485} 208

\bibitem{SS} Sahni V and Shtanov Y, \textit{Braneworld models of dark energy 
}2003\textit{\ J. Cosmol. Astroparticle Phys.} JCAP \textbf{03 }(11) 014

\bibitem{MMT} Maeda K, Mizuno S and Torii T, \textit{Effective Gravitational
Equations on Brane World with Induced Gravity} 2003 \textit{Phys. Rev. }D 
\textbf{68} 024033

\bibitem{Induced} Gergely L \'{A} and Maartens R, \textit{Asymmetric
brane-worlds with induced gravity} 2005 \textit{Phys. Rev.} D \textbf{71}
024032

\bibitem{BraneLuminosityDistance1} Keresztes Z, Gergely L \'{A}, Nagy B and
Szab\'{o} Gy M, \textit{The luminosity-redshift relation in brane-worlds: I.
Analytical results} 2007. In Eq. (15) $\cal{H}$ should read $r$

\bibitem{Goobar} Fairbairn M and Goobar A, \textit{Supernova limits on brane
world cosmology} 2005 \textit{astro-ph/0511029}

\bibitem{Sahni} Alam U and Sahni V, \textit{Confronting Braneworld Cosmology
with Supernova data and Baryon Oscillations} 2006 \textit{Phys. Rev.} D 
\textbf{73} 084024

\bibitem{Sahni02} Alam U and Sahni V, \textit{Supernova Constraints on
Braneworld Dark Energy} 2002 \textit{astro-ph/0209443 }

\bibitem{MM} Maartens R and Majerotto E, \textit{Observational constraints
on self-accelerating cosmology }2006 \textit{Phys. Rev.} D\ \textbf{74}
023004

\bibitem{RoyStrukt} Koyama K and Maartens R, \textit{Structure formation in
the DGP cosmological model} 2006 \textit{J. Cosmol. Astroparticle Phys.}
JCAP \textbf{06 }(01) 016

\bibitem{gao} Barger V, Gao Y, Marfatia D, \textit{Accelerating cosmologies
tested by distance measures} 2007 \textit{Phys. Let.} B \textbf{648} 127

\bibitem{gold} Ries A G, Strolger L-G, Tonry J et al, \textit{Type Ia
Supernova Discoveries at }$z>1$\textit{\ from the Hubble Space Telescope:
Evidence for Past Deceleration and Constraints on Dark Energy Evolution}
2005 \textit{Asrophys. J}. \textbf{607} 665

\bibitem{SNLS} Astier P, Guy J, Regnault J N et al., \textit{The Supernova
Legacy Survey: Measurement of }$\Omega _{M}$\textit{, }$\Omega _{\Lambda }$%
\textit{\ and }$w$\textit{\ from the First Year Data Set} 2006 \textit{%
Astron. Astrophys.} \textbf{447} 31

\bibitem{LMM} Lazkoz R, Maartens R and Majerotto E, \textit{Observational
constraints on phantom-like braneworld cosmologies }2006 \textit{Phys. Rev.}
D \textbf{74} 083510

\bibitem{DabrowskiBrane} D\c{a}browski M P, God\l owski W and Szyd\l owski
M, \textit{Brane universes tested against astronomical data}\ 2004 \textit{%
Int. J. Mod. Phys.} D \textbf{13} 1669

\bibitem{Fay} Fay S, \textit{Branes: cosmological surprise and observational
deception} 2006 \textit{Astron. Astrophys.} \textbf{452} 781

\bibitem{tabletop} Long J C et al., \textit{New experimental limits on
macroscopic forces below 100 microns} 2003 \textit{Nature} \textbf{421} 922

\bibitem{GK} Gergely L \'{A} and Keresztes Z, \textit{Irradiated asymmetric
Friedmann branes} 2006\textit{\ J. Cosmol. Astroparticle Phys.} JCAP \textbf{%
06 }(01) 022

\bibitem{GM} Germani C and Maartens R, \textit{Stars in the braneworld} 2001 
\textit{Phys. Rev. }D \textbf{64} 124010

\bibitem{nucleosynthesis} Maartens R, Wands D, Bassett B A and Heard I P C, 
\textit{Chaotic Inflation on the brane} 2000 \textit{Phys. Rev. }D \textbf{62%
} 041301(R)

\bibitem{BBNLIM} Ichiki K, Yahiro M, Kajino T, Orito M and Mathews G J, 
\textit{Observational Constraints on Dark Radiation in Brane Cosmology} 2002 
\textit{Phys. Rev. }\textbf{D 66} 043521

\bibitem{Nucleosynthesys} Bratt J D, Gault A C, Scherrer R J and Walker T P, 
\textit{Big Bang Nucleosynthesis Constraints on Brane Cosmologies} 2002 
\textit{Phys. Lett}. B \textbf{546} 19

\bibitem{Millennium} Springel V, White S D M, Jenkins A et al, \textit{%
Simulating the joint evolution of quasars, galaxies and their large-scale
distribution} 2005 Nature \textbf{435} 629

\bibitem{ChKN} A. Chamblin, A. Karch and A. Nayeri, \textit{Thermal
Equilibration of Brane-Worlds }2001 \textit{Phys. Lett }B \textbf{509} 163

\bibitem{LSR} Langlois D, Sorbo L and Rodr\'{\i}guez-Mart\'{\i}nez M, 
\textit{Cosmology of a brane radiating gravitons into the extra dimension}
2002 \textit{Phys. Rev. Lett}{. }\textbf{89} 171301

\bibitem{PalStructure} Pal S, \textit{Structure formation on the brane: A
mimicry} 2006 \textit{Phys. Rev.} D \textbf{74} 024005

\bibitem{radiatingBrane} Hebecker A and March-Russell J, \textit{%
Randall--Sundrum II cosmology, AdS/CFT, and the bulk black hole} 2001 
\textit{Nuclear Physics} B \textbf{608} 375

Leeper E, Maartens R and Sopuerta C, \textit{Dynamics of radiating
braneworlds} 2004 \textit{Class. Quant. Grav.} \textbf{21} 1125

Gergely L \'{A}, Leeper E and Maartens R, \textit{Asymmetric radiating
brane-world} 2004 {\ }\textit{Phys. Rev.} D \textbf{70} 104025

Jennings D and Vernon I R, \textit{Graviton emmission into non-Z2 symmetric
brane world spacetimes} 2005 \textit{J. Cosmol. Astroparticle Phys.} JCAP 
\textbf{05 }(07) 011

Langlois D, \textit{Is our Universe brany?} 2006 \textit{Progress of
Theoretical Physics Supplement} \textbf{163} 258

\bibitem{Riess} Riess A G, Filippenko A V, Challis P et al., \textit{%
Observational evidence from Supernovae for an Accelerating Universe and a
Cosmological Constant} 1998 \textit{Astrophys. J. }\textbf{116} 1009

\bibitem{Schmidt} Schmidt B, 2005 in: \textit{The new cosmology,}
Proceedings of the 16th Int. Summer School 2003 Singapore World Scientific

\bibitem{Tonry} Tonry J L, Schmidt B P, Barris B et al., \textit{%
Cosmological Results from High-z Supernovae} 2003 \textit{Astrophys. J.} 
\textbf{594} 1

\bibitem{gold06} Riess A G, Strolger L-G, Casertano S et al., \textit{New
Hubble Space Telescope Discoveries of Type Ia Supernovae at }$z>1$\textit{:
Narrowing Constraints on the Early Behavior of Dark Energy} 2007 to appear
in Astrophys. J. \textbf{656} \textit{astro-ph/0611572}

\bibitem{Garcia} Garcia-Bellido J, \textit{Cosmology and Astrophysics }2005 
\textit{astro-ph/0502139}

\bibitem{KolbTurner} Kolb E W and Turner M S, \textit{The Early Universe}
1990 Adisson Wesley

\bibitem{Tao} Tao C, \textit{Evidence for physics beyond LCDM?} Talk given
at the Cosmology Workshop: Cosmology and Astroparticles at Universit\'{e}
Montpellier 2 France November 23$^{rd}$-24$^{th}$ 2006

\bibitem{Perivolaropoulos} Nesseris S, Perivolaropoulos L, \textit{Tension
and Systematics in the Gold06 SnIa Dataset} 2006 \textit{astro-ph/0612653}
\end{thebibliography}
\end{document}